\documentclass[10pt,a4paper,dvipdfmx]{article}
\usepackage{mathrsfs}
\usepackage{fancyhdr}
\usepackage[top=30truemm,bottom=30truemm,left=20truemm,right=50truemm]{geometry}
\usepackage{indentfirst}
\usepackage{authblk}
\usepackage{graphicx}
\usepackage{bm,color}

\renewenvironment{abstract}{\begin{quotation}{\normalsize }}{\end{quotation}}
\makeatletter
\newcommand{\email}[1]{\newcommand{\@email}{E-mail: #1}}
\renewcommand{\maketitle}{
\newpage\null
    \vspace{2em}
 {\LARGE\bfseries\noindent\ignorespaces\@title\par}
 \vspace{1em}%
 {\large\noindent\ignorespaces\@author\par}
 \vspace{2mm}
 {\normalsize\noindent\ignorespaces\@email\par}
\vspace{1em}
}

\renewcommand\@author{\ifx\AB@affillist\AB@empty\AB@author\else
      \ifnum\value{affil}>\value{Maxaffil}\def\rlap##1{##1}%
     \vspace{1mm} \AB@authlist\\\AB@affillist
    \else  \AB@authors\fi\fi}
\makeatother
\usepackage{amsmath,amssymb}
\title{Multinucleon transfer reaction in time-dependent Hartree-Fock theory}
\author[1,2]{Kazuyuki Sekizawa}
\author[2,3]{Kazuhiro Yabana}
\affil[1]{Faculty of Physics, Warsaw University of Technology, ulica Koszykowa 75, 00-662 Warsaw, Poland}
\affil[2]{Graduate School of Pure and Applied Sciences, University of Tsukuba, Tsukuba 305-8571, Japan}
\affil[3]{Center for Computational Sciences, University of Tsukuba, Tsukuba 305-8577, Japan}
\email{sekizawa@if.pw.edu.pl, yabana@nucl.ph.tsukuba.ac.jp}
\begin{document}
\maketitle

\begin{abstract}
{\bf Abstract: }
Time-dependent Hartree-Fock (TDHF) theory has achieved a remarkable
success in describing and understanding nuclear many-body dynamics
from nucleons' degrees of freedom. We here report our investigation of
multinucleon transfer (MNT) processes employing the TDHF theory.
To calculate transfer probabilities for channels specified by the number
of protons and neutrons included in reaction products, a particle-number
projection (PNP) method has been developed. The PNP method is also
used to calculate excitation energies of reaction products. Combined
use of the PNP method with a statistical model, we can evaluate
MNT cross sections taking account of effects of particle evaporation.
Using these methods, we evaluate MNT cross sections for $^{40,48}$Ca+$^{124}$Sn,
$^{40}$Ca+$^{208}$Pb, and $^{58}$Ni+$^{208}$Pb reactions.
From systematic analyses, we find that cross sections for channels with a 
large reaction probability are in good agreement with experimental data. 
However, the agreement becomes less accurate as the number of transferred 
nucleons increases. Possible directions to improve the description are discussed.
\end{abstract}

{\bf keywords:} TDHF, particle-number projection, multinucleon transfer reaction

\section{Introduction}

In low-energy heavy ion reactions at around the Coulomb barrier,
a variety of phenomena reflecting quantum many-body dynamics
are observed. In subbarrier reactions, fusion and transfer as well 
as quasi-elastic processes take place. The subbarrier fusion process
can be regarded as quantum many-body tunneling phenomena
between two colliding nuclei. In reactions close to the barrier,
a neck is formed between them, providing a pathway of nucleon
exchange between the projectile and the target nuclei. The nucleon
exchange causes an energy dissipation, a transfer of energy from
relative motion of the projectile and the target nuclei into their
internal excitations. While fusion reactions are commonly observed
in low-energy reactions, they are hindered in heavy systems by
the appearance of quasifission processes. Reactions involving
superfluid nuclei attract another interest, effects of pairing 
correlations on nuclear dynamics.

We have been working on multinucleon transfer (MNT) processes,
one of phenomena that attracted substantial interests in low-energy
heavy ion reactions. The MNT processes become dominant in reactions
at energies around and above the Coulomb barrier. Besides curiosity
in their reaction mechanisms, the MNT processes have been attracting
interests as a new means to produce unstable nuclei whose production
is difficult by other methods \cite{Dasso1,Dasso2,TDHF-review,Zag-IQF,
Zag-light,DNS,GRAZING_F}. For example, it has been discussed to use
the MNT reactions to produce neutron-rich nuclei around the neutron
magic number $N=126$ \cite{Zagrebaev(2011),Watanabe(2015)},
which are considered to play a significant role in producing the third
peak structure in the solar abundance.

In the past, theoretical investigations on MNT processes have
been extensively made using semiclassical models such as
{\small GRAZING} \cite{GRAZING}, complex WKB (CWKB)
\cite{CWKB}, a dynamical model based on Langevin-type
equations of motion \cite{Langevin1,Langevin2}, and dinuclear
system model (DNS) \cite{Antonenko(1995),Adamian(1997)1,
Adamian(1997)2,Adamian(1998)}. The semiclassical models of
{\small GRAZING} and CWKB have been quite successful
to describe MNT processes in peripheral collisions (for a review,
see Ref.~\cite{MNT-review}). However, it has turned out that
{\small GRAZING} substantially underestimates cross sections
associated with damped collisions forming reaction products
which have largely different masses from initial ones
\cite{GRAZING_F,MNT-USA}. This is natural because reactions
at a small impact parameter region inside the nuclear rainbow
are not considered in the {\small GRAZING} calculations. The
dynamical Langevin model and DNS have been successful in
describing not only peripheral collisions but also damped collisions
such as deep inelastic collisions, quasifission, fusion-fission, and
fusion reactions. Although the dynamical Langevin model is
recognized as one of the most promising methods to examine
MNT processes including a damped collision, the model contains
some empirical parameters \cite{Zag-IQF} that restrict their
predictive power. To further extend our understanding on
underlying reaction mechanisms and to improve reliability
to predict cross sections, it is highly desired to develop a
microscopic framework for the MNT reaction. To this end,
we have investigated MNT processes employing the
time-dependent Hartree-Fock (TDHF) theory.

As described in detail in other chapters, the TDHF theory has
been quite successful to study nuclear dynamics. Starting
from 1970s, substantial efforts have been devoted for developing
computational and analytical methods, and for applications
(for a review, see Refs.~\cite{TDHF-review,Negele(review)}).
It has been considered that the TDHF theory is suitable to describe
quantities associated with one-body operators \cite{TDHF-review},
and has been applied to averaged quantities in most applications.
To describe MNT cross sections, we go beyond it, extracting transfer
probabilities from the TDHF wave function after collision for each 
channel specified by the nucleon numbers in the reaction products.

In 1977, S.E.~Koonin~$et$~$al.$~proposed a method to calculate
transfer probabilities from a TDHF wave function after collision
\cite{Koonin1977}. In 1983, the same method was applied to
electron transfer processes in atomic collisions by H.J.~L\"ude
and R.M.~Dreizler \cite{Method1983}. The method was also
applied to a collision of a highly-charged ion and an atom 
\cite{Nagano(2000)}. Although the method was useful to
investigate transfer processes, applications were limited to 
relatively small systems because of computational limitations.
The method requires calculations of determinants of $N$-dimensional 
matrices many times, where $N$ stands for the number of particles 
in the system. If we calculate probabilities for all possible processes, 
the number of calculations of the determinants becomes $2^N$.
Thus the method can be applied to systems less than a few
tens of nucleons. In 2011, C.~Simenel proposed an efficient
method \cite{Projection} to calculate the transfer probabilities
using the particle-number projection (PNP) method.
The PNP method succeeds to reduce the computational costs 
significantly, while being analytically equivalent to the method
proposed in the past. The number of calculations of determinants 
necessary to carry out the PNP is about at most several hundreds 
and is independent of the number of particles in the system. 
Using the PNP method, it has become possible to calculate the 
transfer probabilities for reactions involving heavy nuclei.

In this chapter, we will report results of our TDHF calculations
\cite{KS_KY_MNT} for the MNT reactions of
$^{40}$Ca+$^{124}$Sn at $E_{\rm lab}=170$~MeV,
$^{48}$Ca+$^{124}$Sn at $E_{\rm lab}=174$~MeV,
$^{40}$Ca+$^{208}$Pb at $E_{\rm lab}=235$ and 249~MeV, and
$^{58}$Ni+$^{208}$Pb at $E_{\rm lab}=328.4$~MeV,
for which precise experimental data are available
\cite{Corradi(40Ca+124Sn),Corradi(48Ca+124Sn),
Corradi(58Ni+208Pb),Szilner(40Ca+208Pb)2}. We consider
that comparisons between theory and measurements for
those systems with different $N/Z$ ratios and different
charge product $Z_{\rm P} Z_{\rm T}$ will provide
useful information on reaction dynamics \cite{KS_KY_MNT}.

It has been well known that a fast charge equilibration
process takes place in transfer reactions at low-energy where
strength of the driving force for the charge equilibration process
is related to the difference of $N/Z$ ratio between projectile
and target nuclei. While the $^{48}$Ca+$^{124}$Sn system
has a small $N/Z$ asymmetry, all other systems have relatively
large $N/Z$ asymmetries. It has also been well known that
a product of charges of projectile and target nuclei,
$Z_{\rm P}Z_{\rm T}$, determines qualitative aspects of
low-energy heavy ion reactions. When the charge product
$Z_{\rm P}Z_{\rm T}$ exceeds a critical value of about 1600,
occurrence of fusion reactions is substantially suppressed.
The four systems examined have values of
1000 for $^{40,48}$Ca+$^{124}$Sn,
1640 for $^{40}$Ca+$^{208}$Pb, and
2296 for $^{58}$Ni+$^{208}$Pb. 
Since the values are below, around, and above the critical
value, respectively, we expect that different reaction
dynamics will be seen in these systems.

Fragment nuclei produced after MNT processes are expected
to be highly excited. Measurements are carried out for fragments
which are deexcited by emissions of particles and photons.
To compare calculated cross sections with measurements,
proper treatments of deexcitation processes such as nucleon 
evaporation and transfer induced fission are required. Because
these secondary processes take place in a much longer timescale
than that can be simulated by direct time evolution calculations, 
we should estimate the number of nucleons to be evaporated from 
an excited reaction product by a method other than the direct
TDHF calculations. We will use a statistical model for that purpose. 
As an input for such statistical calculations, excitation energy of
a reaction product in each transfer channel is required. Recently, 
we invented a method to calculate the excitation energy using 
the PNP method \cite{KS_KY_PNP}. We review our theoretical 
framework, present calculated cross sections including the particle 
evaporation effect, and compare them with experimental data.

This chapter is organized as follows.
In Sec.~\ref{Sec:Formalism}, we present theoretical framework
and numerical setting of our approach putting emphasis on the
PNP method applied for TDHF wave functions. 
In Sec.~\ref{Sec:Results}, we show results of our TDHF
calculations for MNT processes in $^{40,48}$Ca+$^{124}$Sn,
$^{40}$Ca+$^{208}$Pb, and $^{58}$Ni+$^{208}$Pb
reactions, for which precise experimental data are available.
Effects of particle evaporation on the cross sections are
investigated employing a simple statistical model.
In Sec.~\ref{Sec:Summary}, a summary and a future
prospect are presented.

\section{Theoretical formalism}{\label{Sec:Formalism}}

\subsection{TDHF wave function for heavy ion reactions}

We start with considering a TDHF wave function for low-energy
heavy ion reactions. We denote the number of nucleons of the
projectile and target nuclei as $N_{\rm A}$ and $N_{\rm B}$,
respectively. The total number of nucleons is expressed as
$N=N_ {\rm A}+N_{\rm B}$. In TDHF calculations, the
many-body wave function of the system is expressed by
a single Slater determinant,
\begin{equation}
\Phi({\bf r}_1\sigma_1, \cdots, {\bf r}_N\sigma_N, t)
= \frac{1}{\sqrt{N!}} \det\bigl\{ \phi_i({\bf r}_j\sigma_j, t) \bigr\},
\end{equation}
where $\phi_i({\bf r}\sigma, t)$ denotes single-particle 
wave functions of nucleons where $\sigma$ denotes the spin
coordinate. In this Section, we omit isospin degree of freedom
for simplicity.

The initial TDHF wave function before the reaction is prepared
as a product of two Slater determinants which are the static 
Hartree-Fock (HF) solutions of projectile and target nuclei,
boosted with velocities determined by the initial condition of
relative motion. We divide the space into two subspaces referring
to the position of two nuclei and define number operators for the
two subspaces, $\hat{N}_{\rm A}$ and $\hat{N}_{\rm B}$, by
\begin{equation}
\hat{N}_\tau = \int  \Theta_\tau({\bf r})
\sum_{i=1}^N \delta({\bf r} - {\bf r}_i) \, d{\bf r}
= \sum_{i=1}^N \Theta_\tau({\bf r}_i),
\end{equation}
where $\Theta_\tau({\bf r})$ denotes a space division function defined by
\begin{equation}
\Theta_\tau({\bf r}) = \left\{
\begin{array}{cc}
0 & \mbox{for ${\bf r}\in \tau$}, \\
1 & \mbox{for ${\bf r}\notin \tau$}.
\end{array}\right.
\end{equation}
The initial TDHF wave function is apparently an eigenstate of $\hat{N}_\tau$,
\begin{equation}
\hat{N}_\tau \big|\Phi(t=0)\bigr> = N_\tau \big|\Phi(t=0)\bigr>,
\end{equation}
where $\tau$ refers to a subspace either $A$ or $B$.
Here and hereafter, we often use a bracket notation, $e.g.$
$\big|\Phi(t)\bigr> \equiv \Phi({\bf r}_1\sigma_1,\cdots,
{\bf r}_N\sigma_N, t)$, to simplify notations.

Evolving orbital wave functions by solving the TDHF equation, 
two nuclei approach to each other. Each single-particle
wave function, which initially belongs to either projectile or
target nucleus, starts to extend into the other nucleus.
We consider cases where binary fragments, a projectile-like 
fragment (PLF) and a target-like fragment (TLF), are produced 
after the reaction. At a certain time $t=t_{\rm f}$ at which
the binary reaction products are spatially separated, we again
divide the space into two, $A'$ that includes the PLF and 
$B'$ that includes the TLF. We define the number operators
for the subspaces $A'$ and $B'$ as $\hat{N}_{\rm A'}$ and
$\hat{N}_{\rm B'}$, respectively. The TDHF wave function
after collision is generally not an eigenstate of those operators:
\begin{equation}
\hat{N}_\tau \big|\Phi(t=t_{\rm f})\bigr> \neq N_\tau \big|\Phi(t=t_{\rm f})\bigr>,
\end{equation}
where $\tau=A'$ or $B'$. 

In the following, we denote the TDHF wave function after collision 
at time $t=t_{\rm f}$ as $\big|\Psi\bigr> \equiv \big|\Phi(t=t_{\rm f})\bigr>$
omitting the time index. We also denote single-particle wave functions 
at $t=t_f$ as $\psi_i({\bf r}\sigma)$. We omit a prime from
$A'$ and $B'$ without confusion.

\subsection{Particle-number projection method}

As described above, the TDHF wave function after collision,
$\big| \Psi \bigr>$, is not an eigenstate of the operator
$\hat N_{\tau}$ but a superposition of states with different
mass partitions:
\begin{equation}
\big|\Psi\bigr> = \sum_{n=0}^N \big|\Psi_n\bigr>,
\label{Eq:wf_sum_Psi_n}
\end{equation}
where $\big|\Psi_n\bigr>$ denotes a state in which $n$ nucleons are 
in the spatial region $A$ and $N-n$ nucleons are in the spatial region $B$.

The PNP operator is used to extract a component of a given number of 
nucleons from $\big| \Psi \bigr>$. It is given by \cite{Projection,KS_KY_MNT}
\begin{eqnarray}
\hat{P}_n
&=& \sum_{\{ \tau: A^n B^{N-n} \}}
\Theta_{\tau_1}({\bf r}_1) \cdots \Theta_{\tau_N}({\bf r}_N) \label{Eq:PNPop_SDF}\\
&=& \frac{1}{2\pi}\int_0^{2\pi} e^{i(n - \hat{N}_A)\theta}\, d\theta. \label{Eq:PNPop_Proj}
\end{eqnarray}
The $\{ \tau: A^n B^{N-n} \}$ under the summation in Eq.~(\ref{Eq:PNPop_SDF}) 
means that the sum is taken for all possible combinations where $A$ appears $n$
times and $B$ appears $N-n$ times in the sequence of $\tau_1\tau_2\dots\tau_N$. 

Using the PNP operator, the state $\big|\Psi_n\bigr>$ in Eq.~(\ref{Eq:wf_sum_Psi_n}) 
can be expressed as $\big|\Psi_n\bigr>=\hat{P}_n\big|\Psi\big>$.
The probability $P_n$ that $n$ nucleons are in the spatial region $A$ and
the other $N-n$ nucleons are in the spatial region $B$ is given by
\begin{equation}
P_n = \bigl<\Psi_n\big|\Psi_n\bigr> = \bigl<\Psi\big|\hat{P}_n\big|\Psi\bigr>.
\label{Eq:def_Pn_Proj}
\end{equation}

When we substitute the PNP operator of the form of
Eq.~(\ref{Eq:PNPop_SDF}) into Eq.~(\ref{Eq:def_Pn_Proj}),
we get the formula,
\begin{equation}
P_n = \sum_{\{ \tau: A^n B^{N-n} \}}
\det\bigl\{ \bigl<\psi_i\big|\psi_j\bigr>_{\tau_j} \bigr\},
\label{Eq:Pn_SDF}
\end{equation}
where $\bigl<\psi_i\big|\psi_j\bigr>_\tau \equiv
\sum_\sigma\int d{\bf r}\,\Theta_\tau({\bf r})\,
\psi_i^*({\bf r}\sigma)\psi_j({\bf r}\sigma)$.
To calculate $P_n$ using this formula, determinants
of the matrix $\bigl<\psi_i\big|\psi_j\bigr>_\tau$
need to be calculated as many as $_NC_n$. Calculations
of the determinants become quite difficult once the total
number of nucleons in the system, $N$, becomes large.

When we use the PNP operator of the form of
Eq.~(\ref{Eq:PNPop_Proj}), we get alternative formula given by
\begin{equation}
P_n = \frac{1}{2\pi}\int_0^{2\pi} e^{in\theta} \det\mathcal{B}(\theta)\, d\theta,
\label{Eq:Pn_Proj}
\end{equation}
where $\mathcal{B}(\theta)$ denotes an $N$-dimensional matrix,
\begin{equation}
\mathcal{B}_{ij}(\theta)
= \bigl<\psi_i\big|\psi_j\bigr>_B + e^{-i\theta} \bigl<\psi_i\big|\psi_j\bigr>_A.
\end{equation}
The integral over the phase $\theta$ in Eq.~(\ref{Eq:Pn_Proj}) can be
carried out using the trapezoidal rule discretizing the interval into a 
uniform grid. Using this formula, the number of calculations of the
determinants required to obtain $P_n$ is the number of discrete
points $M$ to carry out the integral numerically, and is independent
of the total number of nucleons in the system. We have found that
$M=200$ is sufficient for the systems analyzed in this chapter.

\subsection{Expectation values in particle-number projected states}

In the particle-number projected wave function, $\big|\Psi_n\bigr>$, 
two reaction products are included: a PLF composed of $n$ nucleons 
inside the spatial region $A$ and a TLF composed of $N-n$ nucleons
inside the spatial region $B$. In this subsection, we present a method
to calculate expectation values of operators for one of the reaction
products included in $\big|\Psi_n\bigr>$ \cite{KS_KY_PNP}.

We first introduce a decomposition of operators according to the
spatial region in which the operator acts. For local one-body operators,
$\hat{\mathcal{O}}^{(1)}$, it is rather trivial:
\begin{eqnarray}
\hat{\mathcal{O}}^{(1)}
&=& \sum_{i=1}^N \bigl[ \Theta_A({\bf r}_i) + \Theta_B({\bf r}_i) \bigr]\,
\hat{o}^{(1)}({\bf r}_i \sigma_i) \nonumber\\
&=& \hat{\mathcal{O}}^{(1)}_A + \hat{\mathcal{O}}^{(1)}_B,
\end{eqnarray}
where $\hat{O}^{(1)}_\tau$ denotes a one-body operator which acts 
only when a nucleon is in the spatial region $\tau$. For local two-body
operators, $\hat{\mathcal{O}}^{(2)}$, we decompose as
\begin{eqnarray}
\hat{\mathcal{O}}^{(2)}
&=& \sum_{i<j}^N
\bigl[ \Theta_A({\bf r}_i) + \Theta_B({\bf r}_i) \bigr]
\bigl[ \Theta_A({\bf r}_j) + \Theta_B({\bf r}_j) \bigr]\,
\hat{o}^{(2)}({\bf r}_i \sigma_i, {\bf r}_j \sigma_j) \nonumber\\
&=& \sum_{i<j}^N
\bigl[ \Theta_A({\bf r}_i)\Theta_A({\bf r}_j) + \Theta_B({\bf r}_i)\Theta_B({\bf r}_j) \nonumber\\[-2mm]
&&\hspace{5mm} + \Theta_A({\bf r}_i)\Theta_B({\bf r}_j) + \Theta_B({\bf r}_i)\Theta_A({\bf r}_j)
\bigr]\, \hat{o}^{(2)}({\bf r}_i \sigma_i, {\bf r}_j \sigma_j) \nonumber\\[1mm]
&=& \hat{\mathcal{O}}^{(2)}_A + \hat{\mathcal{O}}^{(2)}_B + \hat{\mathcal{O}}^{(2)}_{AB},
\end{eqnarray}
where $\hat{\mathcal{O}}^{(2)}_\tau$ denotes a two-body
operator which acts only when both two nucleons are in the spatial
region $\tau$, while $\hat{\mathcal{O}}^{(2)}_{AB}$ denotes
a two-body operator which acts when a nucleon is in the spatial
region $A$ and the other nucleon is in the spatial region $B$.
We note that, when the PLF and the TLF are separated by more
than a distance of nuclear force, only long-ranged Coulomb interaction 
between protons contribute to $\hat{\mathcal{O}}^{(2)}_{AB}$.

The expectation value of the operator $\hat{\mathcal{O}}$
for the reaction product which is composed of $n$ nucleons
and locates inside the spatial region $A$ is given by
\begin{equation}
\mathcal{O}_n^A = \frac{\bigl<\Psi_n\big|\hat{\mathcal{O}}_A\big|\Psi_n\bigr>}{\bigl<\Psi_n\big|\Psi_n\bigr>}.
\label{Eq:def_On_A}
\end{equation}
Using the PNP operator of the form of Eq.~(\ref{Eq:PNPop_Proj}),
we can evaluate the expectation value of Eq.~(\ref{Eq:def_On_A}) as
\begin{equation}
\mathcal{O}_n^A = \frac{1}{2\pi P_n} \int_0^{2\pi} e^{in\theta} \det\mathcal{B}(\theta)\, 
\bigl<\Psi\big|\hat{\mathcal{O}}_A\big|\tilde{\Psi}(\theta)\bigr>\, d\theta,
\label{Eq:On_A_Proj}
\end{equation}
where $\big|\tilde{\Psi}(\theta)\bigr>$ denotes a Slater determinant
composed of the following single-particle wave functions,
\begin{equation}
\tilde{\psi}_i({\bf r}\sigma, \theta) =
\sum_{k=1}^N \psi_k({\bf r}\sigma) \bigl( \mathcal{B}^{-1}(\theta) \bigr)_{ki}.
\end{equation}
We note that, because $\bigl\{ \tilde{\psi}_i(\theta) \bigr\}$
are bi-orthonormal to $\bigl\{ \psi_i \bigr\}$, $i.e.$ $\bigl<
\psi_i\big|\tilde{\psi}_j(\theta)\bigr>=\delta_{ij}$,
calculations of matrix elements, $\bigl<\Psi\big|\hat
{\mathcal{O}}_A \big|\tilde{\Psi}(\theta)\bigr>$, can be
carried out easily. In practical calculations, we need to
perform the PNP for both neutrons and protons. 

We use Eq.~(\ref{Eq:On_A_Proj}) to evaluate the excitation energy
of a reaction product for every transfer channel. As an operator,
we consider the Hamiltonian for the reaction product inside the
spatial region $A$, $\hat{H}_A = \hat{T}_A + \hat{V}_A^{\rm Skyrme}$.
Putting it into Eq.~(\ref{Eq:On_A_Proj}) (more precisely,
using the corresponding formula of the Skyrme energy density
functional kernel), we evaluate the energy expectation value
of the reaction product composed of $N$ neutrons and $Z$
protons, which we denote as $\mathcal{E}_{N,Z}^A$.
To remove the energy associated with a translational motion, 
we make a Galilean transformation for the single-particle wave
functions, moving to the rest frame of the reaction product inside
the spatial region $A$. The excitation energy of the reaction product
is evaluated by subtracting the ground state energy,
\begin{equation}
E_{N,Z}^*(E,b) \equiv \mathcal{E}_{N,Z}^A(E,b) - E_{N,Z}^{\rm g.s.}.
\label{Eex}
\end{equation}

\subsection{MNT cross sections with evaporation effects}

For a given incident energy $E$ and impact parameter $b$, 
we have a TDHF wave function after collision, $\Psi(E,b)$. 
We can calculate the probability for a production of a nucleus 
having $N$ neutrons, $P_N(E,b)$, and that having $Z$ protons, 
$P_Z(E,b)$, using Eq.~(\ref{Eq:Pn_Proj}). The probability producing
the reaction product having $N$ neutrons and $Z$ protons is given
by $P_{N,Z}(E,b) = P_N(E,b) P_Z(E,b)$. Integrating it over the
impact parameter, we obtain the corresponding cross section,
\begin{equation}
\sigma_{NZ} (E)= 2\pi \int_{b_{\rm min}}^{\infty} b\, P_{N,Z}(E,b)\, db,
\end{equation}
where $b_{\rm min}$ denotes the minimum impact parameter
at which we observed binary reaction products after collision.

The above formula gives us the MNT cross sections in which the
numbers of neutrons and protons correspond to those immediately
after the fragments are formed. Reaction products of MNT processes
immediately after the separation are highly excited, especially for
collisions at small impact parameters. Therefore, a proper evaluation
of effects of secondary disintegration processes is necessary. Since
these processes take place in a much longer timescale than the
timescale producing binary fragment nuclei, we treat the effect
by a statistical model.

As a first investigation \cite{KS_KY_Evap}, we have taken a simple
statistical model of particle evaporation developed by I.~Dostrovsky
and his coworkers \cite{Dostrovsky}. In this model, evaporation of 
neutrons, protons, deuterons, tritons, $^3$He, and $^4$He 
are taken into account. As an input, this model only requires
the excitation energy of the nucleus, $E^*$. Using this model,
we evaluate evaporation probabilities, $P^{n,z}_{N,Z}[E^*_{N,Z}]$, 
emitting $n$ neutrons and $z$ protons from the nucleus composed
originally of $N$ neutrons and $Z$ protons with excitation energy $E^*_{N,Z}$. 
The MNT cross sections incorporating the evaporation effects are then given by
\begin{equation}
\sigma_{N,Z}(E) = 2\pi \int_{b_{\rm min}}^\infty
b\, \sum_{n,z} P_{N+n,Z+z}(E,b) P^{n,z}_{N+n,Z+z}\bigl[ E_{N+n,Z+z}^*(E,b) \bigr]\, db,
\end{equation}
where the summation over $n$ and $z$ is taken for all possible
numbers of evaporated nucleons. Excitation energy $E_{N+n,Z+z}^*(E,b)$
is calculated by employing Eq.~(\ref{Eex}).

\subsection{Numerical details}{\label{Sec:Numerical_details}}

We have developed our own TDHF code for heavy ion reactions
\cite{KS_PhD}. In the code, single-particle wave functions are
represented on a uniform grid in three-dimensional Cartesian
coordinates without any symmetry restrictions. We set the mesh
spacing to be 0.8~fm. We use 11-point finite-difference formula
for the first and second derivatives. For time evolution calculations,
we utilize the 4th-order Taylor expansion method in operating the
time propagation operator. Coulomb potential is calculated employing
the Hockney's method \cite{Hockney} where Fourier transformations
are performed in a box with the side length two times larger than the
simulation box to obtain the potential of isolated boundary condition. 
To prepare the ground state, we used $26 \times 26 \times 26$ grid
points. For the time evolution calculations, we used $60 \times 60 \times 26$
grid points. For the PNP calculation, we use the trapezoidal rule with
$M$ discrete points. We set $M=200$ for all calculations presented below.
Calculations are carried out for impact parameter region up to 10~fm.

\section{Results}{\label{Sec:Results}}

We use the Skyrme energy density functional, SLy5 parameter set. 
To obtain ground states, we first perform constrained Hartree-Fock
calculations with deformation parameters $\beta=0$, 0.1, and 0.2,
and $\gamma=0^\circ$, $30^\circ$, and $60^\circ$. We then
remove these constraints and re-minimize the energy. Among
solutions, we regard the minimum energy solution as the Hartree-Fock
ground state. The ground states of $^{40,48}$Ca and $^{208}$Pb
are spherical, while $^{124}$Sn and $^{58}$Ni are found to be
oblate and prolate shapes, respectively, with $\beta \sim 0.11$
for both systems. In collision calculations, the symmetry axis of
these deformed nuclei is set perpendicular to the reaction plane
of our TDHF calculations. In Ref.~\cite{KS_KY_MNT}, we showed
that reaction processes do not depend much on the direction of
the  deformation for $^{40}$Ca+$^{124}$Sn reaction.

The relative motion between the PLF and the TLF after the collision 
is characterized by the deflection angle $\Theta$ and the total 
kinetic energy loss (TKEL). They are calculated in the following way.
At a certain time $t=t_{\rm f}$ after the collision, we consider
two spheres around the center-of-mass of two reaction products,
${\bf R}_{\tau}(t_{\rm f})$ ($\tau=A$ or $B$), with a radius
10~fm for the PLF and 14~fm for the TLF. Integrating densities
of neutrons and protons inside the spheres, we evaluate the
average number of nucleons, $A_\tau(t_{\rm f})$, and protons
$Z_\tau(t_{\rm f})$. The reduced mass of the reaction products
is given by $\mu(t_{\rm f})=M_A(t_{\rm f})M_B(t_{\rm f})/
(M_A(t_{\rm f})+M_B(t_{\rm f}))$, where $M_\tau =A_\tau m$
and $m$ is the nucleon mass. Using the relative vector ${\bf R}(t)
={\bf R}_A(t)-{\bf R}_B(t)$ and its time derivative $\dot{\bf R}(t)
=\bigl( {\bf R} (t+\Delta t)-{\bf R}(t-\Delta t) \bigr)/(2\Delta t)$, 
we evaluate the total kinetic energy (TKE) of outgoing fragments
at infinity which is given by,
\begin{equation}
{\rm TKE} = \frac{1}{2}\mu(t_{\rm f})\dot{\bf R}^2(t_{\rm f})
+ \frac{Z_A(t_{\rm f})Z_B(t_{\rm f})e^2}{\bigl|{\bf R}(t_{\rm f})\bigr|}.
\end{equation}
The TKEL is given by ${\rm TKEL}=E_{\rm rel}-{\rm TKE}$,
where $E_{\rm rel}$ is the incident relative energy. The deflection
angle $\Theta$ can be extracted from ${\bf R}(t_{\rm f})$ and
$\dot{\bf R}(t_{\rm f})$ assuming that relative motion continues
along the Coulomb trajectory for $t>t_{\rm f}$.

\begin{figure}[t]
\begin{center}
\includegraphics[width=14cm]{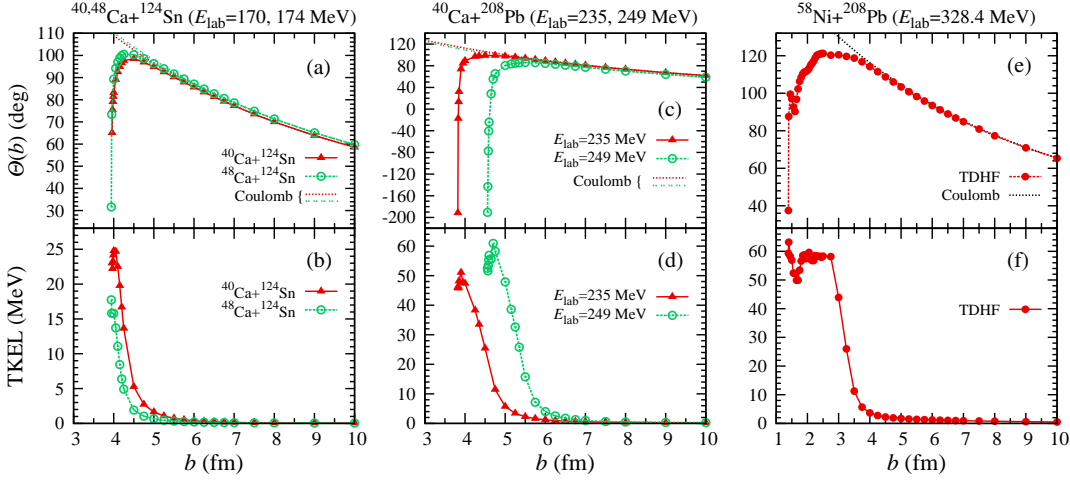}
\caption{
Deflection angle and total kinetic energy loss
as functions of the impact parameter for
$^{40,48}$Ca+$^{124}$Sn (a,~b),
$^{40}$Ca+$^{208}$Pb (c,~d),
$^{58}$Ni+$^{208}$Pb (e,~f).
Data were taken from Ref.~\cite{KS_KY_MNT}.
}
\label{FIG:TKEL+Theta}
\end{center}
\end{figure}

In Fig.~\ref{FIG:TKEL+Theta}, we show deflection angle and TKEL 
for $^{40,48}$Ca+$^{124}$Sn (a,~b), $^{40}$Ca+$^{208}$Pb (c,~d), 
and $^{58}$Ni+$^{208}$Pb (e,~f), as functions of the impact parameter. 
When the impact parameter is sufficiently large, the deflection angle
coincides with that of the Coulomb trajectory. As the impact parameter 
decreases, the deflection angle becomes smaller because of the 
nuclear attractive interaction. As shown in Ref.~\cite{KS_KY_MNT}, 
the Coulomb rainbow angles agree reasonably with a peak position 
of measured angular distributions.

\begin{figure}[t]
\begin{center}
\includegraphics[width=10cm]{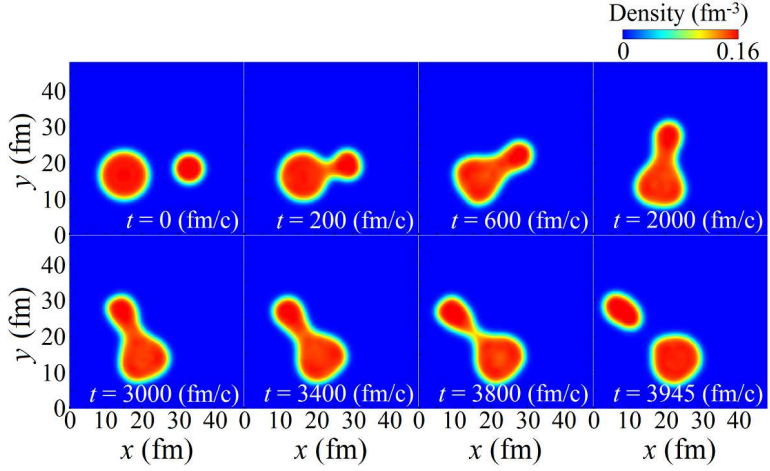}
\caption{
Snapshots of the density on the reaction plane for $^{58}$Ni+$^{208}$Pb
reaction at $E_{\rm lab}=328.4$~MeV and $b=1.39$~fm. The figure
was taken from Ref.~\cite{KS_KY_MNT}.
}
\label{FIG:rho(t)_58Ni+208Pb}
\end{center}
\end{figure}

The TKEL increases as the impact parameter decreases for all
systems, as the deflection angle starts to deviate from the
Coulomb trajectory. The maximum values reach 25~MeV and 18~MeV
for $^{40,48}$Ca+$^{124}$Sn, 50~MeV and 60~MeV for
$^{40}$Ca+$^{208}$Pb at $E_{\rm lab}=235$~MeV and 249~MeV, 
and 60~MeV for $^{58}$Ni+$^{208}$Pb. These values are
correlated with the charge product of projectile and target nuclei,
$Z_{\rm P}Z_{\rm T}$. The relatively small values in
$^{40,48}$Ca+$^{124}$Sn reflect that they are easily fused
when two nuclei touch even slightly. As the charge product
$Z_{\rm P}Z_{\rm T}$ increases, fusion suppression takes
place. At a small impact parameter region, two nuclei collide
more deeply and form a thick neck structure. Since two nuclei
depart after forming the thick neck, produced fragments are
highly excited and the TKEL becomes large for $^{40}$Ca,$
^{58}$Ni+$^{208}$Pb reactions. We note that the TKEL is
almost constant for a certain range of impact parameter for
$^{58}$Ni+$^{208}$Pb, indicating a full momentum transfer
in this system. In Fig.~\ref{FIG:rho(t)_58Ni+208Pb}, we show
snapshots of the density for the $^{58}$Ni+$^{208}$Pb reaction
at $E_{\rm lab} = 328.4$~MeV and $b=1.39$~fm. In this reaction,
a formation of thick neck is seen during the reaction.

\begin{figure}[t]
\begin{center}
\includegraphics[width=10.8cm]{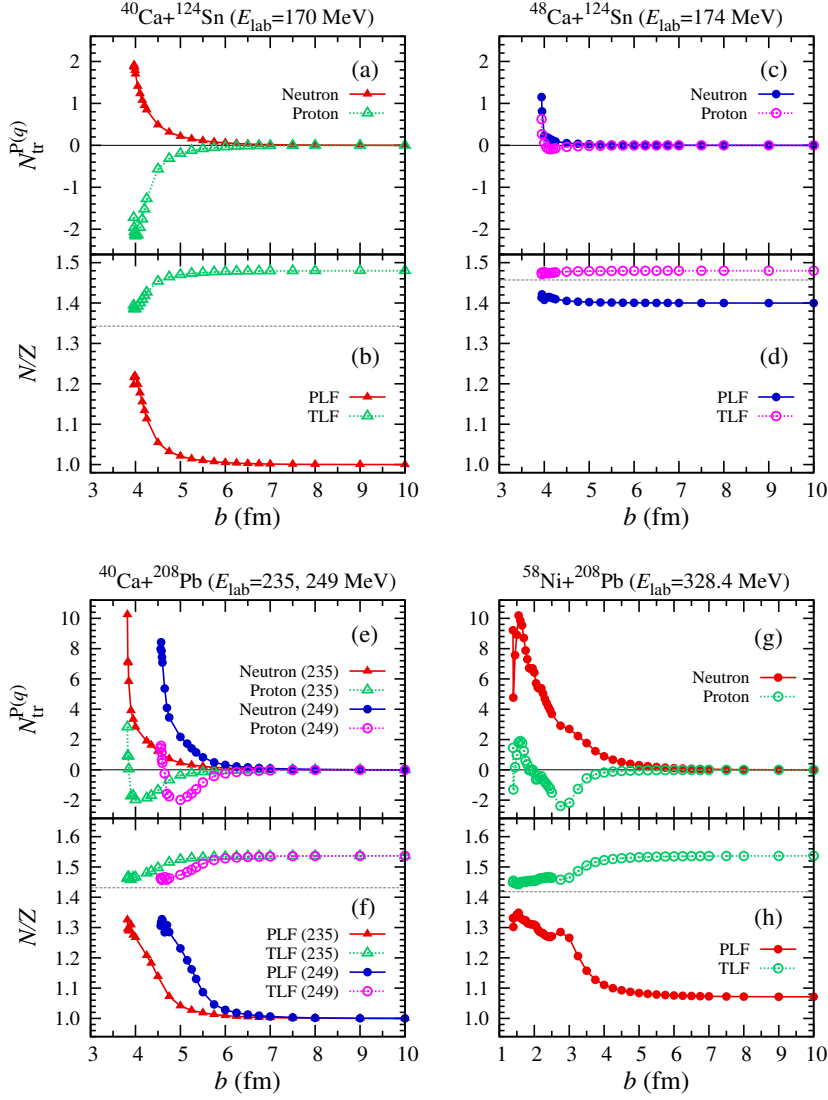}
\caption{
Average number of transferred nucleons and $N/Z$ ratios of
PLF and TLF are shown as functions of the impact parameter for
$^{40}$Ca+$^{124}$Sn (a,~b),
$^{48}$Ca+$^{124}$Sn (c,~d),
$^{40}$Ca+$^{208}$Pb (e,~f),
$^{58}$Ni+$^{208}$Pb (g,~h).
Data were taken from Ref.~\cite{KS_KY_MNT}.
}
\label{FIG:Ntr+NZ-ratio}
\end{center}
\end{figure}

Figure~\ref{FIG:Ntr+NZ-ratio} shows average number
of transferred nucleons and $N/Z$ ratio as functions of 
impact parameter for four systems,
$^{40}$Ca+$^{124}$Sn (a,~b),
$^{48}$Ca+$^{124}$Sn (c,~d),
$^{40}$Ca+$^{208}$Pb (e,~f), and
$^{58}$Ni+$^{208}$Pb (g,~h).
In panels (a,c,e,g), a plus sign indicates that nucleons are
transferred from target to projectile, while a minus sign
indicates opposite. As seen from the panels, we find that
neutrons are always transferred from the target to the
projectile. Protons are transferred opposite, from the projectile
to the target at large impact parameter region. However,
the direction of the proton transfer changes at small impact
parameter region. 

From panels (b,d,f,h), we find that transfers proceed toward
the $N/Z$ equilibrium between the projectile and the target nuclei. 
Since the $^{48}$Ca+$^{124}$Sn is very close to the $N/Z$ equilibrium
from the beginning, a very small number of nucleons are transferred
on average as seen in Fig.~\ref{FIG:Ntr+NZ-ratio}~(c). Other systems,
($^{40}$Ca+$^{124}$Sn and $^{40}$Ca,$^{58}$Ni+$^{208}$Pb),
have a relatively large $N/Z$ asymmetry. The direction of the nucleon
transfer except for protons at small impact parameter is consistent with
the direction expected from the initial $N/Z$ ratios of the projectile and
the target nuclei.

As the charge product $Z_{\rm P}Z_{\rm T}$ increases,
binary fragments are produced even at a small impact parameter,
because of the suppressed fusion reaction. In such collisions,
we find a qualitative change in transfer mechanisms.
In $^{40}$Ca,$^{58}$Ni+$^{208}$Pb reactions, the number
of transferred neutrons increases monotonically, while the number
of transferred protons first increases and then decreases.
At a certain small impact parameter, the number of transferred
protons changes its sign. At a very small impact parameter region,
both neutrons and protons are transferred toward the same direction,
from target to projectile. This change in the transfer behavior is
related to the thick neck formation. When the neck is broken,
nucleons in the neck region are absorbed by one of the fragments.
Because the neck is composed of both neutrons and protons,
the absorption of the nucleons in the neck results in transfer
of neutrons and protons in the same direction.

\begin{figure}[t]
\begin{center}
\includegraphics[width=10.8cm]{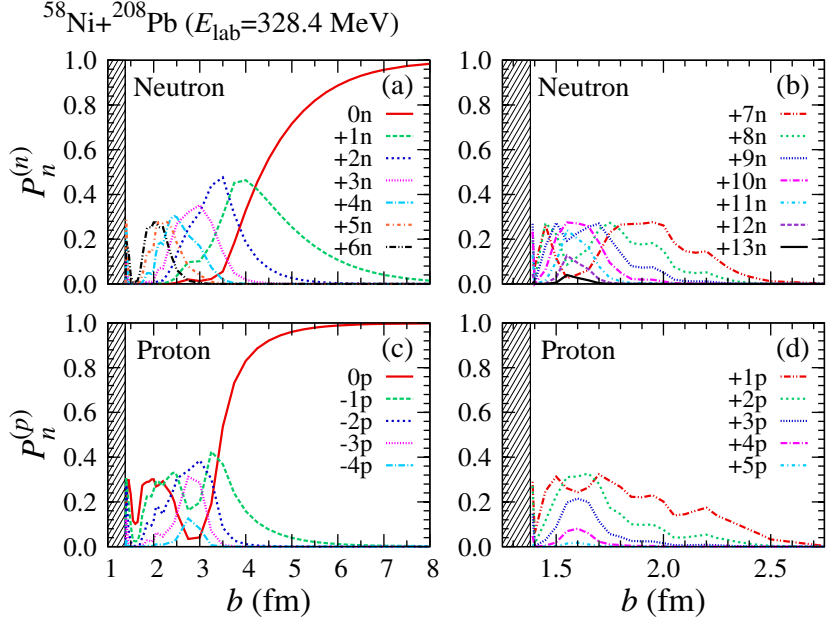}
\caption{
Transfer probabilities as functions of the impact parameter for
$^{58}$Ni+$^{208}$Pb reaction at $E_{\rm lab}=328.4$~MeV.
The figure was taken from Ref.~\cite{KS_KY_MNT}.
}
\label{FIG:Pn_58Ni+208Pb}
\end{center}
\end{figure}

Using the PNP method, we obtain probability for each transfer
channel as a function of the impact parameter from the TDHF 
wave functions after collision. As an example, we show transfer
probabilities for $^{58}$Ni+$^{208}$Pb reaction in
Fig.~\ref{FIG:Pn_58Ni+208Pb}. Panels (a) and (b) show
neutron transfer probabilities, while panels (c) and (d) show
proton transfer probabilities. The shaded region indicates an
impact parameter region in which fusion reactions take place.
We note that the right panels (b) and (d) show results at
a small impact parameter region.

As seen from the figure, one-neutron transfer probability 
($+1$n, green dashed line in (a)) has a long tail up to 8~fm. 
One-proton transfer probability ($-1$p, green dashed line
in (c)) shows somewhat shorter tail compared with that of
neutron. As the impact parameter decreases, transfer
probabilities for various transfer channels become appreciable.
The $+x$n in (a) indicates that $x$ neutrons are transferred from
$^{208}$Pb to $^{58}$Ni. The $-x$p in (c) indicates that
$x$ protons are transferred from $^{58}$Ni to $^{208}$Pb.
The probabilities of neutron transfer in opposite direction are
very small and are not shown. At a impact parameter region
smaller than 3~fm shown in panels (b) and (d), we find
probabilities of many-neutron transfer from $^{208}$Pb to
$^{58}$Ni, up to transfer of 13 neutrons. For protons, we
find probabilities of transfer of up to 5 protons from
$^{208}$Pb to $^{58}$Ni. The direction of transfer of
neutrons and protons are consistent with the observation
in Fig.~\ref{FIG:Ntr+NZ-ratio}~(g). Namely, neutrons and
protons are transferred in opposite directions at a large impact
parameter region, toward the charge equilibrium of the system.
At a small impact parameter region, neutrons and protons
transfer toward the same direction.

\begin{figure}[t]
\begin{center}
\includegraphics[width=10.5cm]{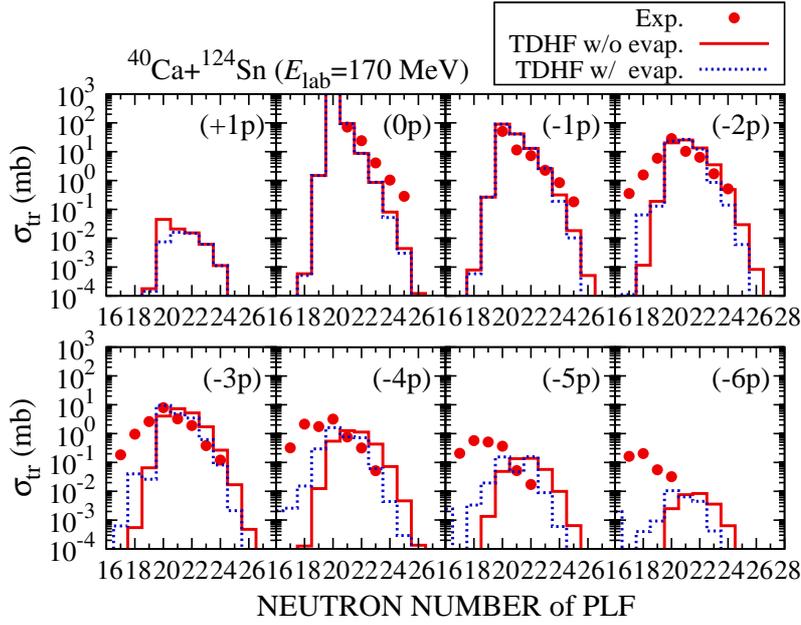}
\caption{
Production cross sections for PLF in $^{40}$Ca+$^{124}$Sn
reaction at $E_{\rm lab}=170$~MeV. The cross sections
are classified according to the number of transferred protons
which is indicated by ($\pm x$p). The horizontal axis is the
number of neutrons in the PLF. Red filled circles denote
experimental data. Red solid lines (blue dotted lines) show
the results of TDHF calculations without (with) effects of
particle evaporation.
}
\label{FIG:NTCS_40Ca+124Sn}
\end{center}
\end{figure}

\begin{figure}[b]
\begin{center}
\includegraphics[width=12cm]{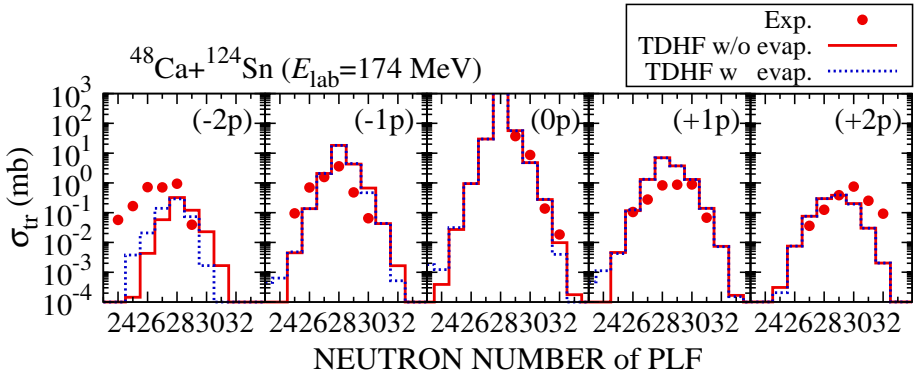}
\caption{
Same as Fig.~\ref{FIG:NTCS_40Ca+124Sn}, but for
$^{48}$Ca+$^{124}$Sn reaction at $E_{\rm lab}=174$~MeV.
}
\label{FIG:NTCS_48Ca+124Sn}
\end{center}
\end{figure}

\begin{figure}[t]
\begin{center}
\includegraphics[width=10.5cm]{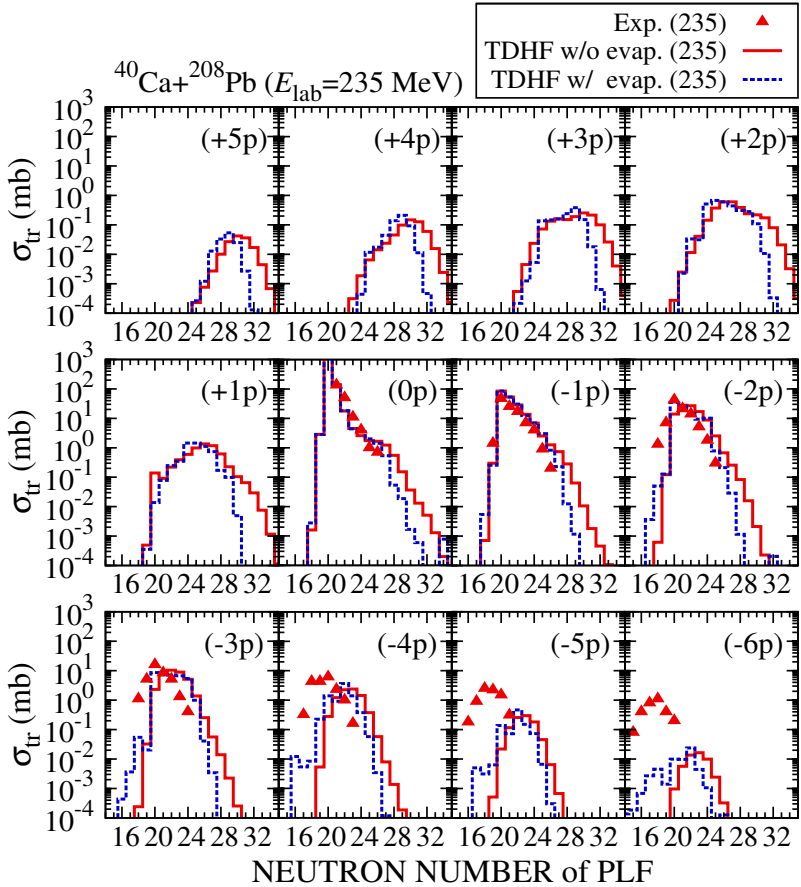}
\caption{
Same as Figs.~\ref{FIG:NTCS_40Ca+124Sn} and
\ref{FIG:NTCS_48Ca+124Sn}, but for
$^{40}$Ca+$^{208}$Pb reaction at $E_{\rm lab}=235$~MeV.
}
\label{FIG:NTCS_40Ca+208Pb_E235}
\end{center}
\end{figure}

\begin{figure}[t]
\begin{center}
\includegraphics[width=10.5cm]{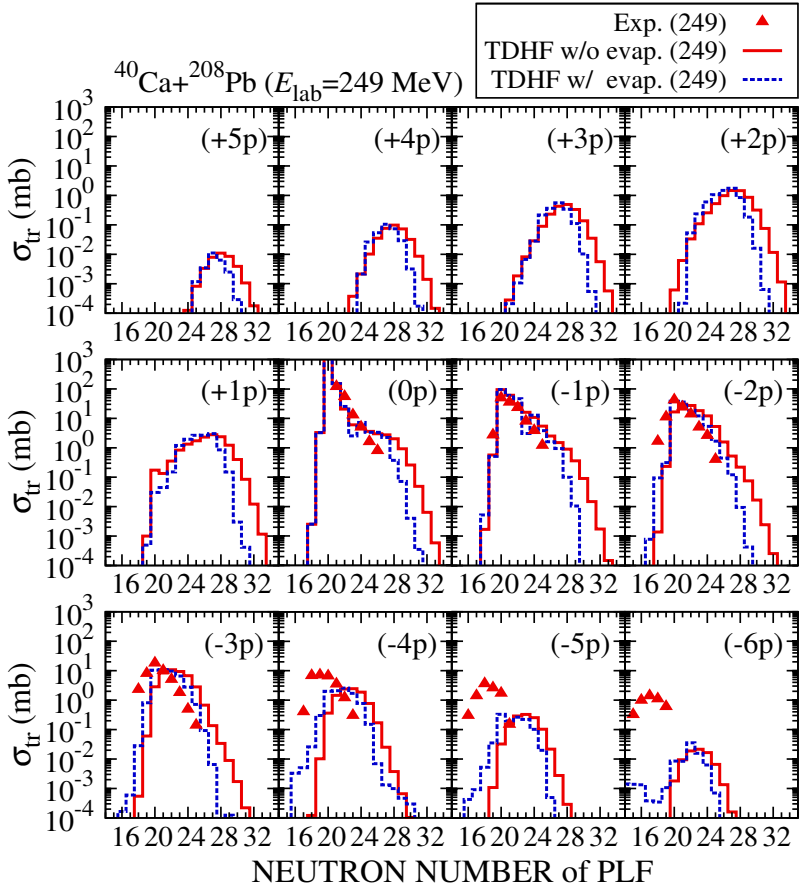}
\caption{
Same as Figs.~\ref{FIG:NTCS_40Ca+124Sn},
\ref{FIG:NTCS_48Ca+124Sn}, and
\ref{FIG:NTCS_40Ca+208Pb_E235}, but for
$^{40}$Ca+$^{208}$Pb reaction at $E_{\rm lab}=249$~MeV.
}
\label{FIG:NTCS_40Ca+208Pb_E249}
\end{center}
\end{figure}

\begin{figure}[t]
\begin{center}
\includegraphics[width=10.5cm]{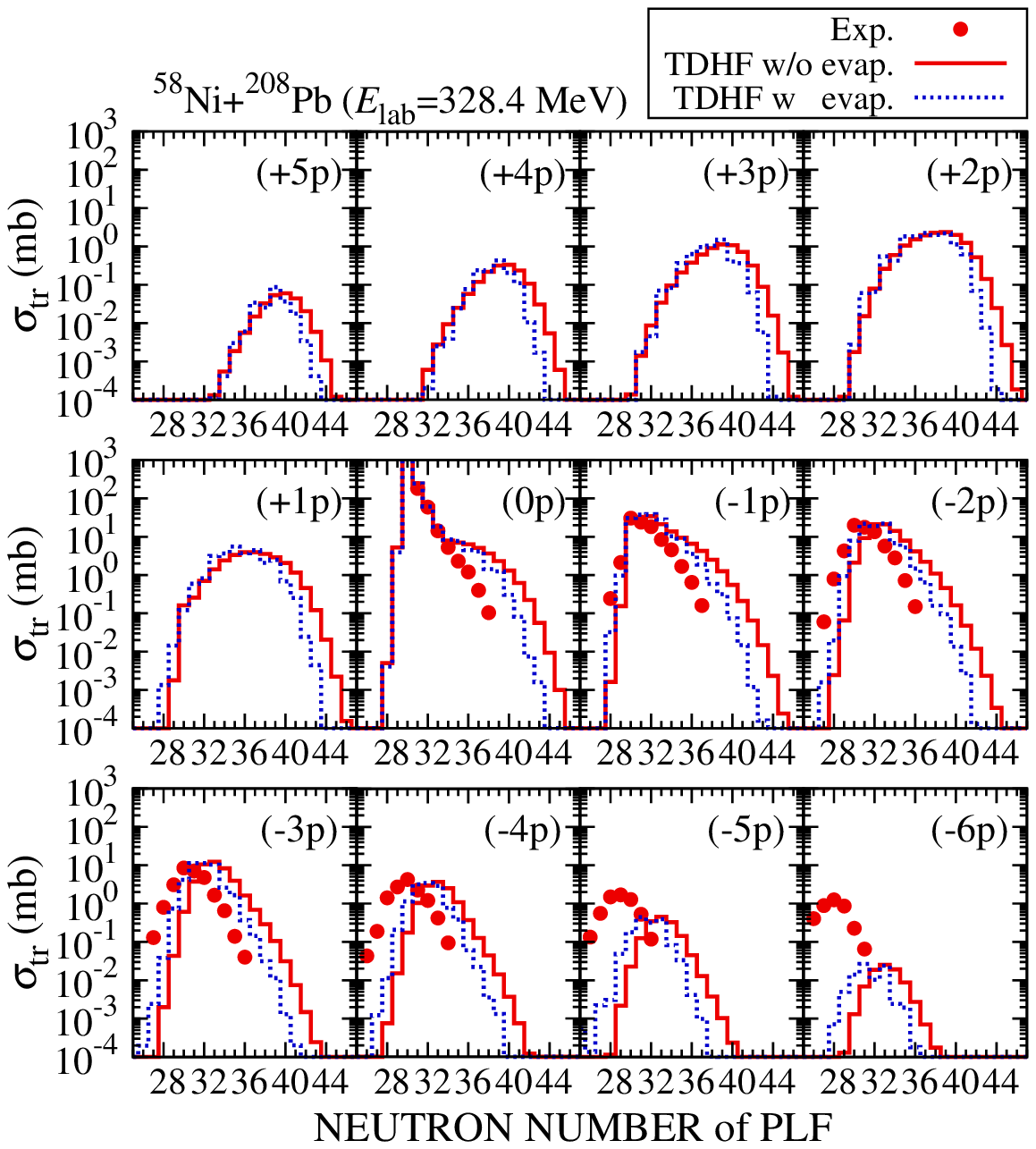}
\caption{
Same as Figs.~\ref{FIG:NTCS_40Ca+124Sn},
\ref{FIG:NTCS_48Ca+124Sn}, \ref{FIG:NTCS_40Ca+208Pb_E235},
and \ref{FIG:NTCS_40Ca+208Pb_E249}, but for
$^{58}$Ni+$^{208}$Pb reaction at $E_{\rm lab}=328.4$~MeV.
}
\label{FIG:NTCS_58Ni+208Pb}
\end{center}
\end{figure}

In Figs.~\ref{FIG:NTCS_40Ca+124Sn} to \ref{FIG:NTCS_58Ni+208Pb},
we show MNT cross sections,
$^{40}$Ca+$^{124}$Sn at $E_{\rm lab}=170$~MeV in Fig.~\ref{FIG:NTCS_40Ca+124Sn},
$^{48}$Ca+$^{124}$Sn at $E_{\rm lab}=174$~MeV in Fig.~\ref{FIG:NTCS_48Ca+124Sn},
$^{40}$Ca+$^{208}$Pb at $E_{\rm lab}=235$~MeV and
249~MeV in Figs.~\ref{FIG:NTCS_40Ca+208Pb_E235} and \ref{FIG:NTCS_40Ca+208Pb_E249}, and
$^{58}$Ni+$^{208}$Pb at $E_{\rm lab}=328.4$~MeV in Fig.~\ref{FIG:NTCS_58Ni+208Pb}.
In all Figures, red points show experimental data, red-solid lines
show calculated cross sections without the evaporation effects,
and blue-dotted lines show calculated cross sections with the
evaporation effects. We first compare calculated cross sections
before taking account of the evaporation effects, given by red-solid
lines, with measured cross sections. The cross sections for channels
with small number of transferred nucleons are reasonably reproduced
by the calculation for all systems. Except for $^{48}$Ca+$^{124}$Sn
reaction where $N/Z$ ratio is similar between the projectile and
the target, cross sections are dominated by the transfer of protons
from lighter to heavier nuclei. As the number of transferred protons
increases, measured cross sections as a function of the neutron
number of the PLF show peaks at less than that of the projectile.
This indicates that both protons and neutrons are transferred in
the same direction. However, in TDHF calculations without taking
account of evaporation effects, the peak position locates at the
neutron number more than that of the projectile. As a possible
origin of the discrepancy, we mentioned the effect of evaporation
from excited fragments \cite{KS_KY_MNT}. 

The cross sections taking account of evaporation effects are shown
by blue-dotted lines. As seen from figures, the discrepancy is
improved to some extent. However, the shift by the evaporation
effect is not sufficient to fully remove the discrepancy. In other
approaches such as the GRAZING, there remain discrepancies
even after the evaporation effects are included, consistent with
the present results. We consider the discrepancy suggests
significance of the correlation effects that require description
beyond the TDHF theory. We note that recently the same analysis
was performed for an asymmetric system, $^{18}$O+$^{206}$Pb,
at an above barrier energy, which shows similar discrepancy
\cite{Bidyut(2015)}.

The limitation of accuracy of the TDHF calculations comes from
the fact that, for a collision specified by the impact parameter
and the energy, there is only one mean-field that is used to
evolve the system. From Fig.~\ref{FIG:Ntr+NZ-ratio}, we see
that we have trajectories in which at most two protons are transferred
($-2$p), on average, from the projectile to the target. From these
solutions, we extract probabilities accompanying up to 6 protons
($-6$p) by the PNP method. When the average number of
transferred protons is about two from the projectile to the
target, we find about two neutrons are transferred ($+2$n)
from the target to the projectile. Because of this, our calculation
results in the cross sections which show a peak at the two neutron
transfer ($+2$n), irrespective of the number of proton transfers.
To further improve the description, we need to employ the theory
which goes beyond the TDHF theory, such as the method of
Balian-V\'en\'eroni \cite{BV(1981),Simenel(BV2011)}, the
stochastic mean-field \cite{SMF(2008),SMF(2009)1,SMF(2009)2,
SMF(2012),SMF(2013),SMF(2014)} model, and the time-dependent
generator coordinate method (TDGCM) \cite{TDGCM1,TDGCM2}.

\section{Summary and prospect}{\label{Sec:Summary}}

In this chapter, we reviewed our recent attempt to
investigate multinucleon transfer (MNT) processes by the 
time-dependent Hartree-Fock (TDHF) theory combined with 
the particle-number projection (PNP) method. The TDHF theory
has been successful to investigate nuclear many-body dynamics.
Owing to a continuous development of computational facilities,
it becomes possible to conduct systematic TDHF calculations
for various systems with different initial conditions. The PNP
method enables us to analyze the TDHF wave function after
collision, providing reaction probabilities for a specific channel
specified by the number of nucleons involved in the fragments.

As typical examples, we showed analyses of MNT processes
in $^{40,48}$Ca+$^{124}$Sn, $^{40}$Ca+$^{208}$Pb,
and $^{58}$Ni+$^{208}$Pb reactions for which precise 
experimental data are available. For all systems examined, 
we have found that nucleons are transferred toward the
direction of the charge equilibrium. For systems with a large
difference of $N/Z$ ratio between the projectile and the target,
($^{40}$Ca+$^{124}$Sn, $^{40}$Ca+$^{208}$Pb, and
$^{58}$Ni+$^{208}$Pb), transfers of a number of nucleons
are observed in the calculation. For a system with a small
$N/Z$ asymmetries, $^{48}$Ca + $^{124}$Sn, nucleons
do not transfer much on average.

As the charge product of the projectile and the target nuclei,
$Z_{\rm P}Z_{\rm T}$, increases, we found that the two
nuclei are once connected by forming a thick neck and then
reseparate. When the thick neck ruptures, nucleons inside
the neck are absorbed by one of the fragments, resulting in
transfers of many neutrons and protons toward the same
direction. We consider that this neck breaking process is
a precursor of quasifission process which dominates in small
impact parameter collisions between heavy nuclei.

For each transfer channel specified by the number of neutrons
and protons included in reaction products, we evaluate transfer
probabilities and excitation energies of the reaction products by
the PNP method. We also evaluate evaporation probabilities of
excited fragments employing a statistical model. Using these
quantities, we calculate production cross sections of each
reaction product including effects of particle evaporation.
The obtained cross sections are compared with measurements.

From the comparisons, we have found that the TDHF theory
nicely reproduces measured cross sections when the number
of transferred nucleons is small. As the number of transferred
nucleons increases, however, we have found that the accuracy
of the TDHF theory to reproduce measured cross sections is
reduced. Although the inclusion of the evaporation effects
has improved the agreement between experimental and
calculated results, the calculations still substantially
underestimate the cross sections when a number of
nucleons are transferred. This failure indicates a necessity
of improving the theoretical framework.

We are now extending our study toward quasifission-induced
MNT processes in a collision between heavy nuclei. In such reactions,
disintegration processes of not only particle evaporation but also
transfer-induced fissions should be considered. Since the PNP method
allows us to evaluate expectation values of any observables of
reaction products for each transfer channel, we plan to use them
in sophisticated statistical models, such as PACE, CASCADE, GEMINI,
or HIVAP, to evaluate effects of secondary processes on measurable
cross sections.

Finally we would like to mention possible directions to improve
theoretical descriptions of MNT processes. A well-known limitation
of the TDHF calculation is the fact that it employs a single mean-field
in the time evolution calculation for a given initial condition, although
we extract reaction probabilities of different numbers in the fragments
from the solution. To include the difference of the mean-field for transfer
channels, the time-dependent generator coordinate method (TDGCM) 
\cite{TDGCM1,TDGCM2} will be useful. Although practical applications
of the TDGCM to heavy ion reactions are challenging, it should be
a promising extension for MNT reactions. We note that the pairing
correlation is not taken into account in the present calculation
employing the TDHF theory. The time-dependent Hartree-Fock-Bogoliubov
(TDHFB) theory will definitely provide useful information to this
direction \cite{Hashimoto(2007),Avez(2008),Stetcu(2011),
Hashimoto(2012),Hashimoto(2013),Bulgac(fission)}.

Recently, a gamma-ray spectroscopic study of reaction products
of MNT reactions has become feasible \cite{gamma_MNT,136Xe+238U(2015)}.
Although we have only achieved number projection in the present work,
our method can be extended to include parity and angular momentum
projections \cite{Ring-Schuck}. Calculations and making comparison
with experimental data will provide us rich information on microscopic
reaction mechanisms.

\section*{Acknowledgments}
We would like to appreciate Prof. J.A.~Maruhn for his prominent
contributions to nuclear physics society, especially for continuous
studies using the time-dependent Hartree-Fock theory. This research
work used computational resources of the HPCI system provided
by Information Initiative Center, Hokkaido University, through the
HPCI System Research Project (Project IDs: hp120204, hp140010,
and hp150010). This work was supported by the Japan Society for
the Promotion of Science (JSPS) Grant-in-Aid for JSPS Fellows,
Grant Number 25-241.

\end{document}